# ***PROJECT REPORT***

## **Submitted by:**

Chiranjeev Kumar
10100EN038
B.Tech (Part III), VI- Semester
Department of Computer Science and Engineering,
Indian Institute of Technology (BHU), Varanasi

## **Supervisor:**

Dr. A.K. TRIPATHI

PROFESSOR,

DEPT. OF COMP. SCI. & ENGG.

INDIAN INSTITUTE OF TECHNOLOGY

(BANARAS HINDU UNIVERSITY),

VARANASI.

# ACKNOWLEDGEMENT

I would like to convey my deepest gratitude to Dr. A.K. TRIPATHI, who guided me through this project. His keen interest, motivation and advice helped me immensely in successfully completing this project.

I would also like to thank Dr. R. B. Mishra, Head of the Department, Computer Science and Engineering, for allowing us to avail all the facilities of the Department necessary for this project.

**Chiranjeev Kumar**

# CERTIFICATE

This is to certify that, Mr. CHIRANJEEV KUMAR has completed his project on "**making a java application which will teach approximation algorithm"** satisfactorily in partial fulfilment under the department of Computer Engineering IIT(BHU),Varanasi during academic year 2012-2013.

__________________________

Teacher In-Charge

# ABSTRACT

This application for learning APPROXIMATION ALGORITHM has been designed in Java which will make user comfortable in learning the very complex subject “NP-Completeness” and the solution to NP-Complete problem using approximation algorithm.

# Table of Contents

# 1. Introduction

There are various problems in computer science for which no polynomial time algorithm is known till now. These special problems are categorized as NP (not polynomial time) problems. A problem P is ***NP-hard*** if a polynomial-time algorithm for P would imply a polynomial-time algorithm for *every problem in NP*.

Many problems of practical significance are NP-complete, yet they are too important to abandon merely because we don't know how to find an optimal solution in polynomial time. Even if a problem is NP-complete, there may be hope. We have at least three ways to get around NP-completeness.

- First, if the actual inputs are small, an algorithm with exponential running time may be perfectly satisfactory.
- Second, we may be able to isolate important special cases that we can solve in polynomial time.
- Third, we might come up with approaches to find *near-optimal* solutions in polynomial time (either in the worst case or the expected case). In practice, near-optimality is often good enough. We call an algorithm that returns near-optimal solutions an ***approximation algorithm***. This project is all about development of polynomial-time approximation algorithms for several NP-complete problems.

# 2. Working of the application

Once the application is started, this will guide the user step-by-step and the user will learn the subject in an easy way.

The starting window looks like-

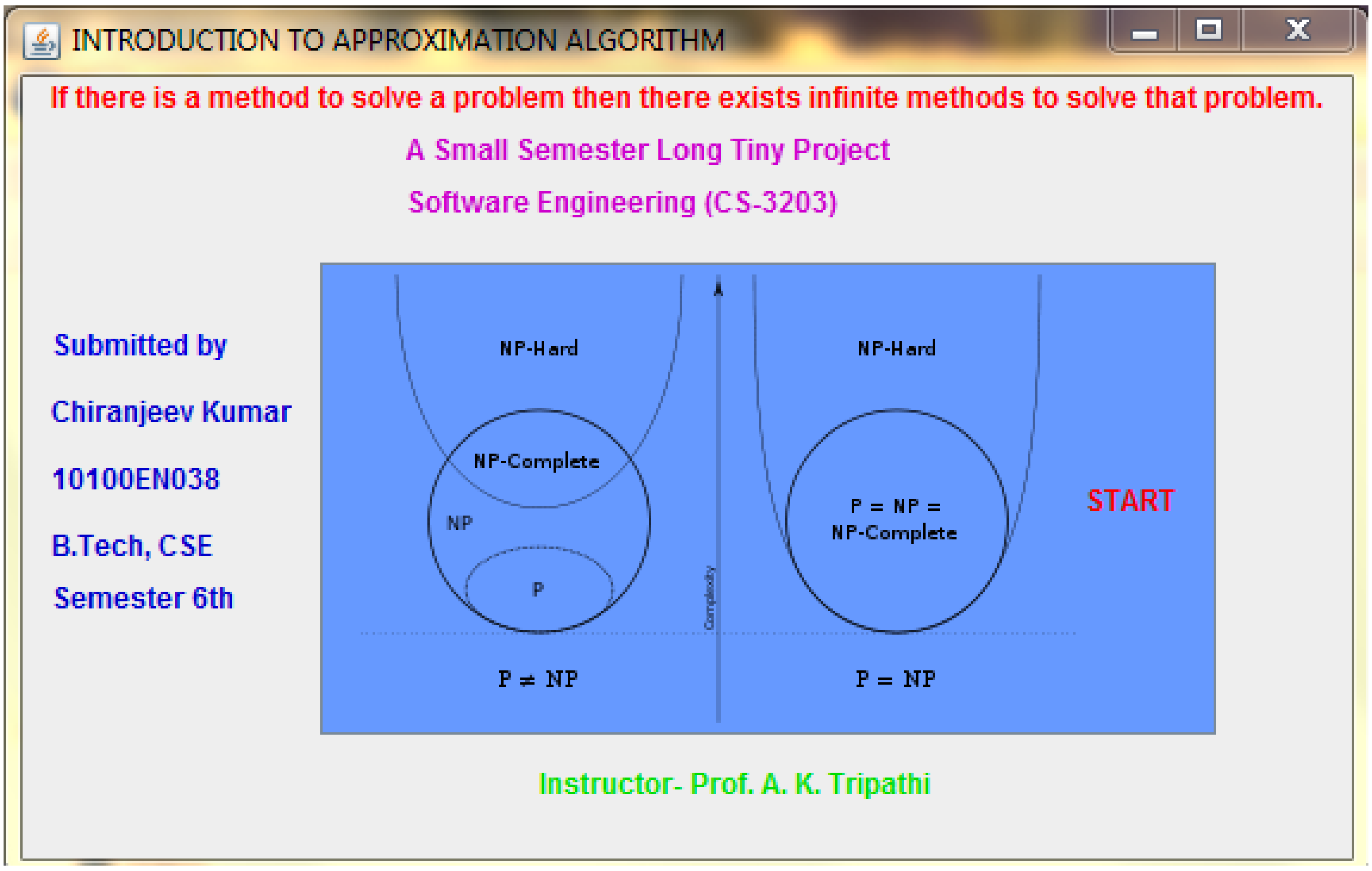

If the user clicks on the rectangular box, next window appears like-

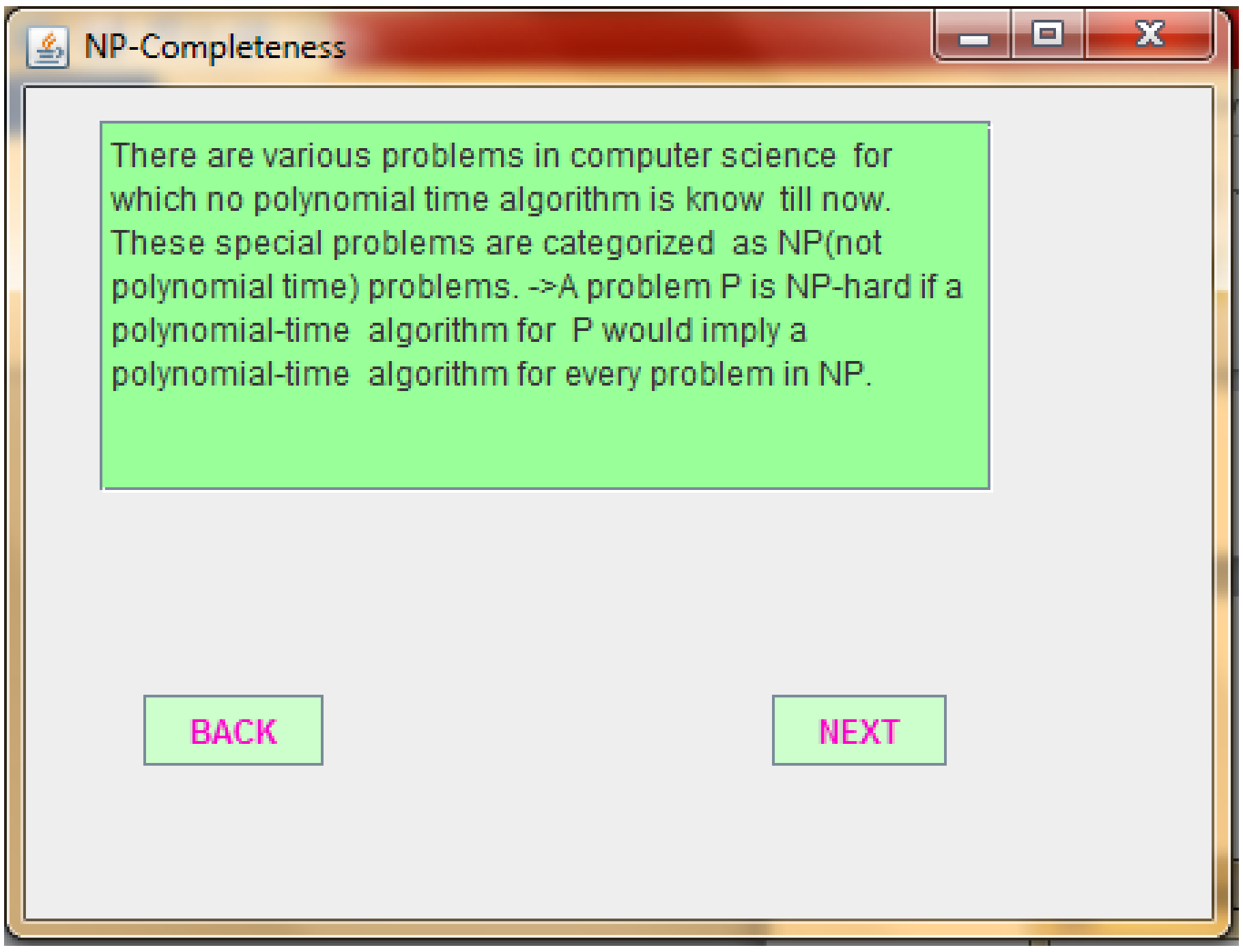

Pressing next button will open the next window like-

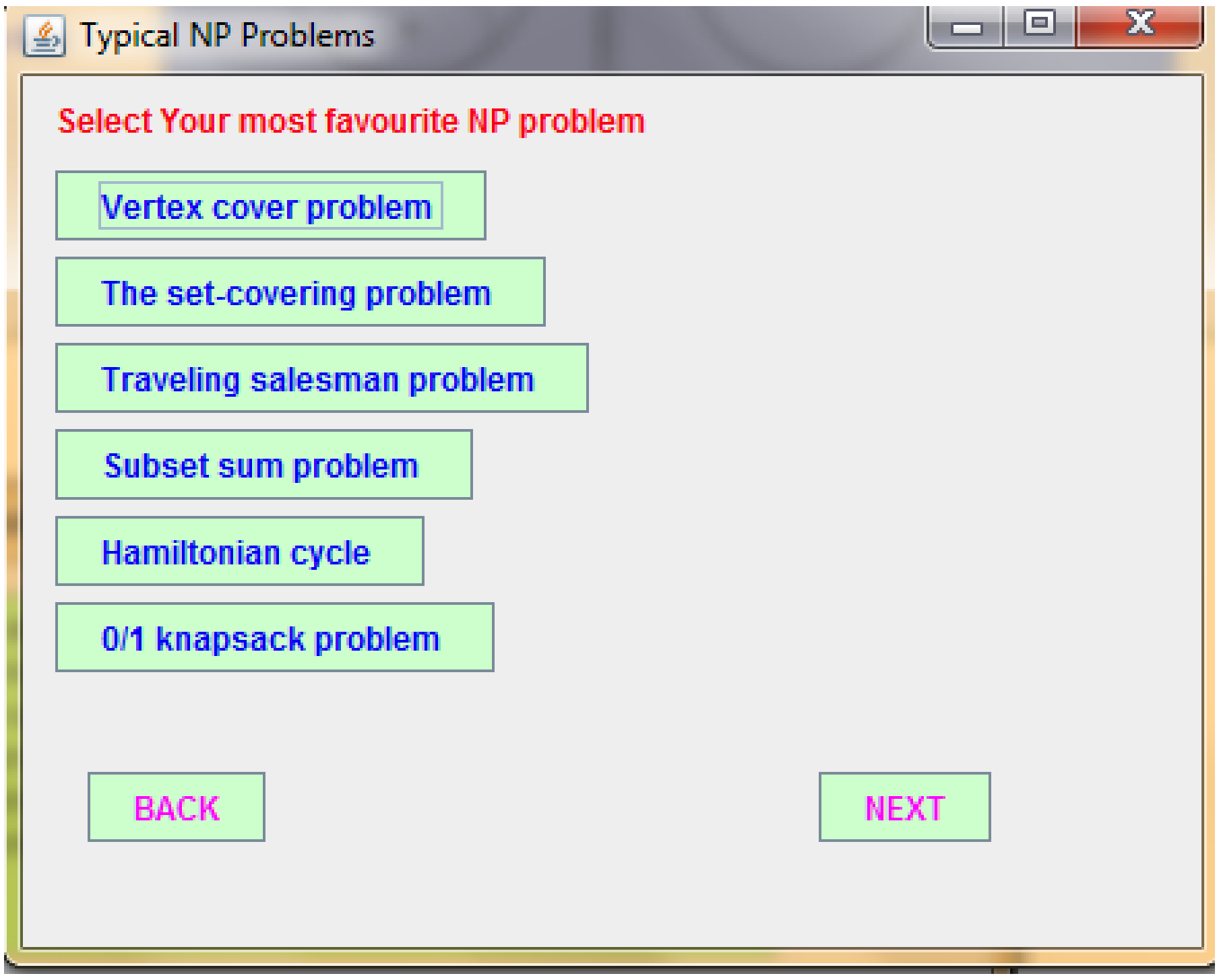

Problem list

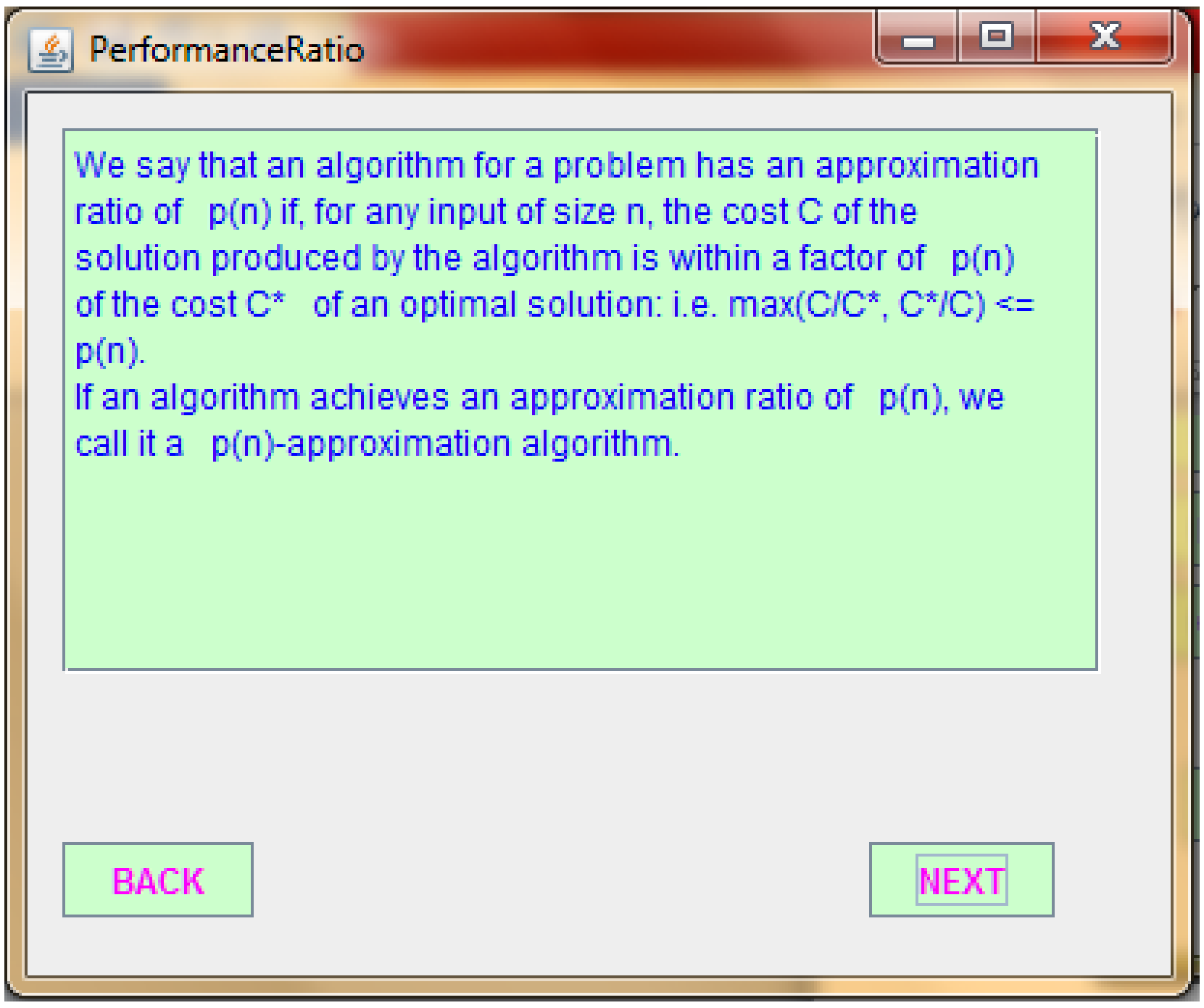

Utilities and concepts

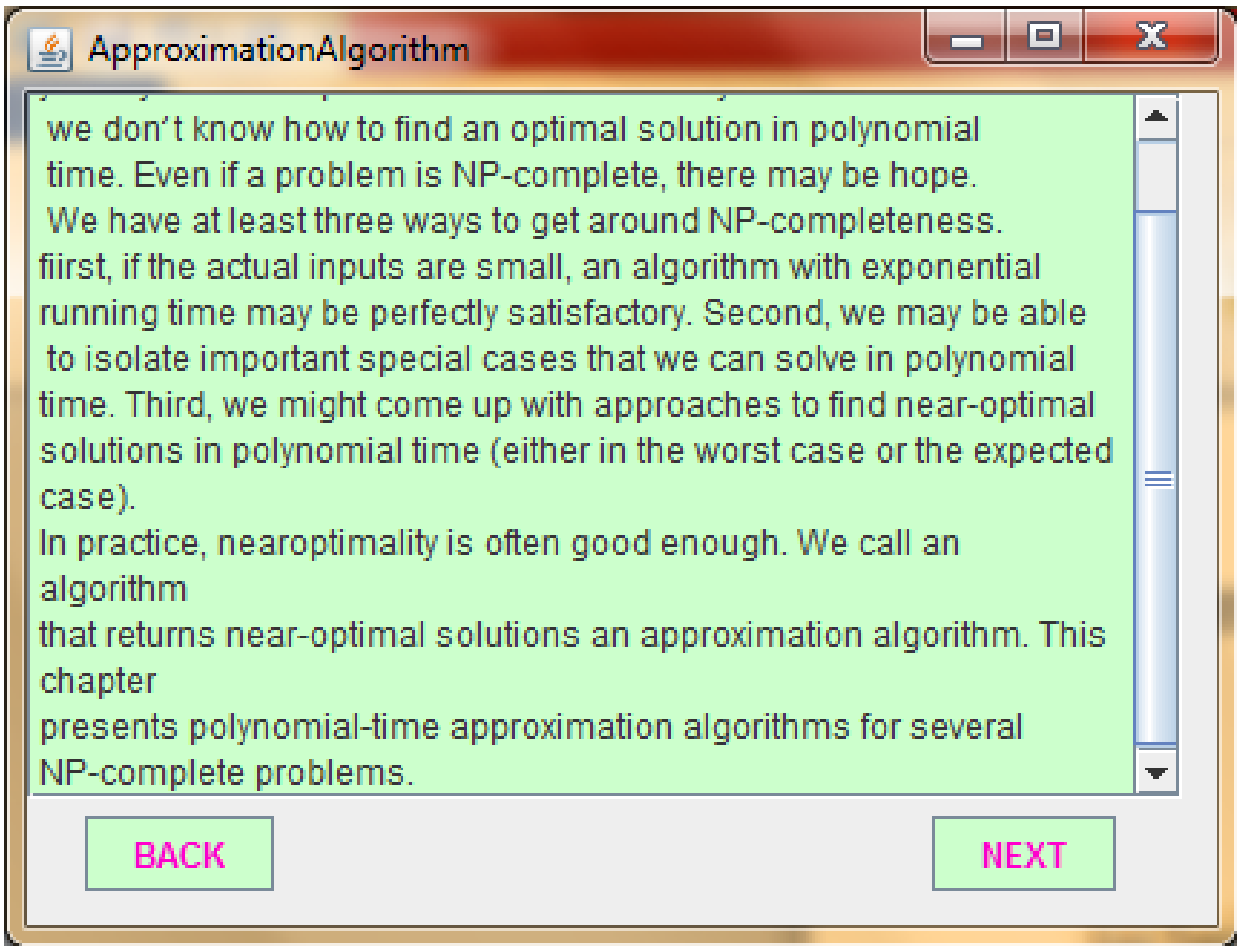

# 3. Uses of the application

An old engineering slogan says, “fast. Cheep. Reliable. Choose two.” Similarly, if P!= NP, we can’t simultaneously have two algorithms that (1) find optimal solution (2) in polynomial time (3) for any instance. At least one of these requirements must be relaxed in any approach to dealing with an NP-hard optimization problem.

One approach relaxes the “for any instance” requirement, and finds the polynomial time algorithms for special cases of the problem at hand. This is useful if the instances one desires to solve fall into one of these special cases, but this is not frequently the case.

A more common approach is to relax the requirement of polynomial-time solvability. The goal is then to find optimal solutions to problems by clever exploration of the full set of possible solutions to a problem. This is often a successful approach if one is willing to take minutes or even hours to find the best possible solution; perhaps even more importantly, one is never certain that for the next input encountered, the algorithm will terminate in any reasonable amount of time. This is the approach taken by those in the field of operations research and mathematical programming who solves integer programming formulations of discrete optimization problems, or those in the area of artificial intelligence who consider techniques such as A* search or constraint programming.

By far the most common approach; however is to relax the requirement of finding an optimal solution, and instead settle for a solution that is “good enough,” especially if it can be found in seconds or less. There has been an enormous study of various types of heuristics and metaheuristics such as simulated annealing, genetic algorithms, and tabu search, to name but a few. These techniques often yield good results in practice.

Any user that wants to learn approximation algorithm and NP-Completeness can use this application in an effective way. This application is specifically for beginner.

# 4. Approximation techniques

No polynomial-time algorithm has yet been discovered for an NP-complete problem, nor has anyone yet been able to prove that no polynomial-time algorithm can exist for any one of them. This so-called P ¤ NP question has been one of the deepest, most perplexing open research problems in theoretical computer science since it was first posed in 1971. Several NP-complete problems are particularly tantalizing because they seem on the surface to be similar to problems that we know how to solve in polynomial time.

**Performance ratios for approximation algorithms-**

We say that an algorithm for a problem has an ***approximation ratio*** of ρ(n) if, for any input of size n, the cost C of the solution produced by the algorithm is within a factor of ρ(n) of the cost C* of an optimal solution:

Max($\frac{C}{C*}$, $\frac{C*}{C}$) <= ρ(n).

If an algorithm achieves an approximation ratio of ρ(n), we call it a ρ(n) **–** ***approximation algorithm.***

Some approximation techniques used in this application are-

1. Greedy Algorithms and local search
2. Rounding Data and Dynamic Programming
3. Deterministic rounding of linear programs
4. Rounding sampling and randomized rounding of linear programs
5. Randomized rounding of semi-definite programs
6. The prime dual method
7. Cuts and metrics
8. Further uses of the techniques

# 5. Approximation methods

Some Problems used in the application have been implemented using approximation algorithms. These solutions are mostly polynomial time 2-approximation algorithms.

1. **Vertex-cover problem**

The ***vertex-cover problem*** is to find a vertex cover of minimum size in a given undirected graph. We call such a vertex cover an ***optimal vertex cover***. This problem is the optimization version of an NP-complete decision problem. Even though we don't know how to find an optimal vertex cover in a graph G in polynomial time, we can efficiently find a vertex cover that is near-optimal. The following approximation algorithm takes as input an undirected graph G and returns a vertex cover whose size is guaranteed to be no more than twice the size of an optimal vertex cover.
It's polynomial time approximation algorithm is:

1. C = Ø
2. E' = G.E
3. **while** E != Ø
4. let (u ,v) be an arbitrary edge of E'
5. C = C U {u ,v}
6. remove from E' every edge incident on either u or v
7. **return** C

2. **Travelling Salesman Problem**

In the traveling-salesman problem, we are given a complete undirected graph G =(V,E) that has a nonnegative integer cost c(u,v) associated with each edge (u,v) ε E, and we must find a hamiltonian cycle (a tour) of G with minimum cost. TSP is an NP-complete problem. We formalize this notion by saying that the cost function c satisfies the ***triangle inequality*** if, for all vertices u,v,w ε V, C(u,v) <= c(u,w) + c(w,v). The traveling-salesman problem is NP-complete even if we require that the cost function satisfy the triangle inequality. Thus, we should not expect to find a polynomial-time algorithm for solving this problem exactly. Instead, we look for good approximation algorithms.

APPROX-TSP-TOUR(G,c)

1. select a vertex r ε G.$V$ to be a “root” vertex
2. compute a minimum spanning tree T for G from root r using MST-PRIM(G, c, r)
3. Let H be a list of vertices, ordered according to when they are first visited in a preorder tree walk of T.
4. **return** the hamiltonian cycle H

- The used approximation algorithm is a greedy approach
- even with a simple implementation of MST-PRIM, the running time of APPROX-TSP-TOUR is , O($V^2$), We now show that if the cost function for an instance of the traveling-salesman problem satisfies the triangle inequality, then APPROX-TSP-TOUR returns a tour whose cost is not more than twice the cost of an optimal tour.
- APPROX-TSP-TOUR is a polynomial-time 2-Approximation algorithm for the traveling-salesman problem with the triangle inequality.

3. **Subset-sum problem**

An instance of the subset sum problem is a pair (S,t), where S is a set {x1,x2,x3….xn} of positive integers. This decision problem asks whether there exists a subset of S that adds up exactly to the target value t. This problem is NP-complete.

- EXACT-SUBSET-SUM(S,t)

1. N = |S|
2. L0 = <0>
3. For i=1 to n
    1. Li = MERGE-LISTS(Li-1, Li-1 +xi )
    2. Remove from Li every element that is greater than t
4. Return the largest element in Ln.

- APPROX-SUBSET-SUM(S, t, ε̇)

1. N = |S|
2. L0 = <0>
3. For i=1 to n
    1. Li = MERGE-LISTS(Li-1, Li-1 +xi )
    2. Li = TRIM(Li, ε̇/2n)
    3. Remove form Li every element that is greater than t
4. Let z* be the largest value in Ln
5. Return z*.

# 6. Implementation

Because this project is somewhat small and straightforward, a waterfall type of model has been used. We start the design activity by identifying classes of objects in the problem domain and relationship between the classes. From the problem specification we can clearly identify the objects as the list of NP-problems as well as their solutions. Besides that one extra main class has been used. All the necessary data structure and utility functions have been separately implemented in different classes to reduce coupling.

# 7. Results

The application is packaged in JAR file. Some windows describing the solution to the problems on executing the application appear as-

Subset-sum problem window

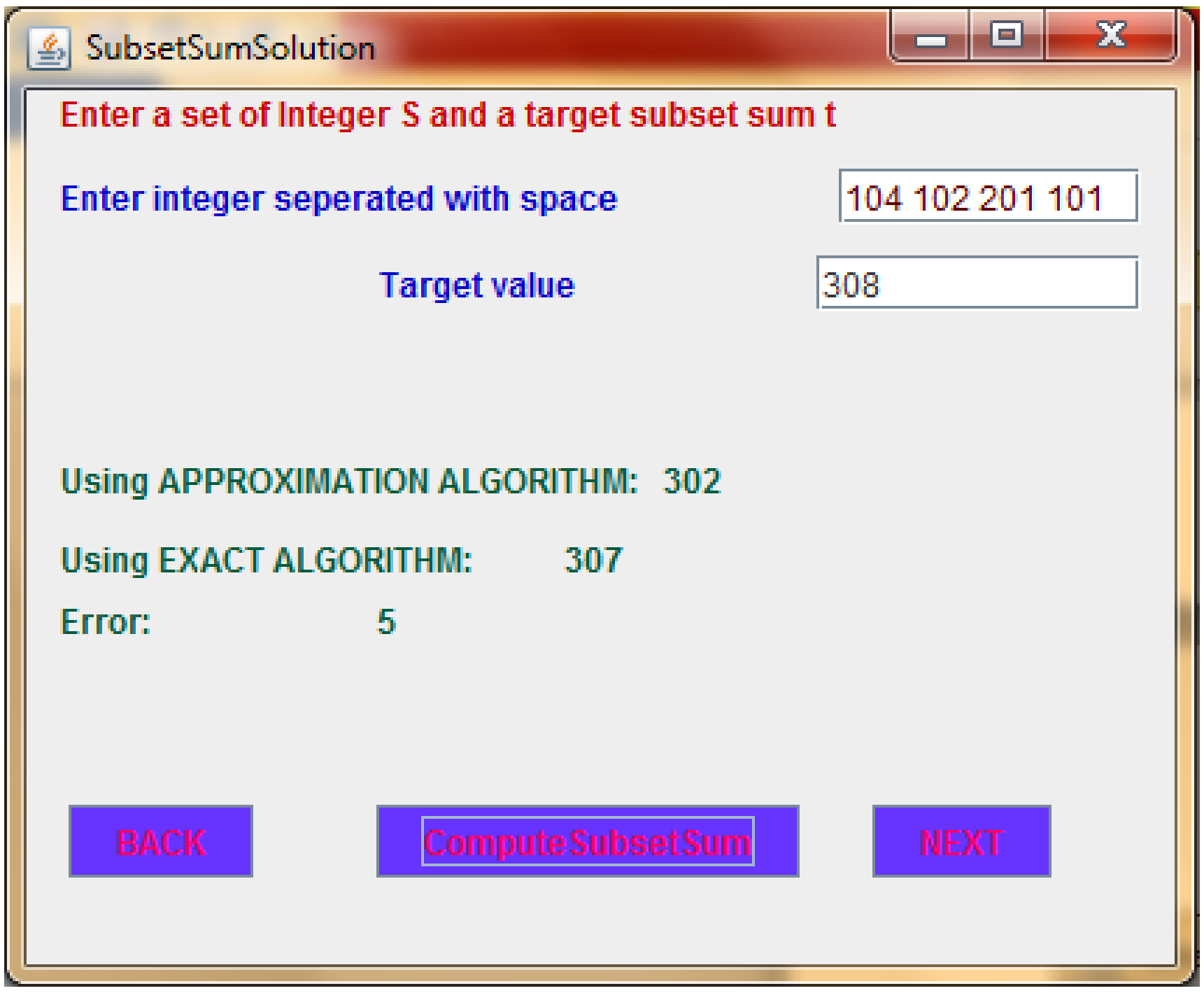

Hamiltonion cycle problem

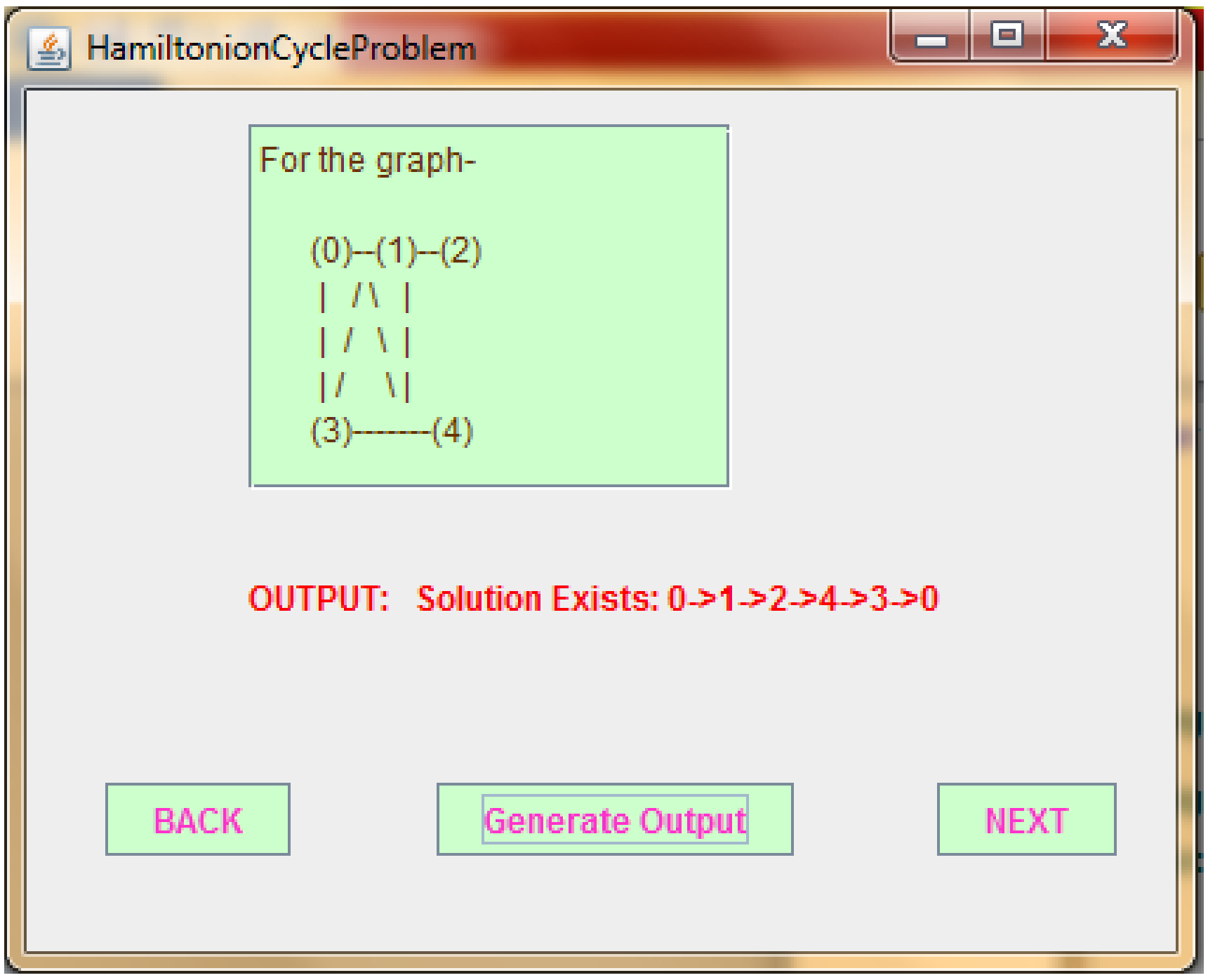

Knapsack problem window

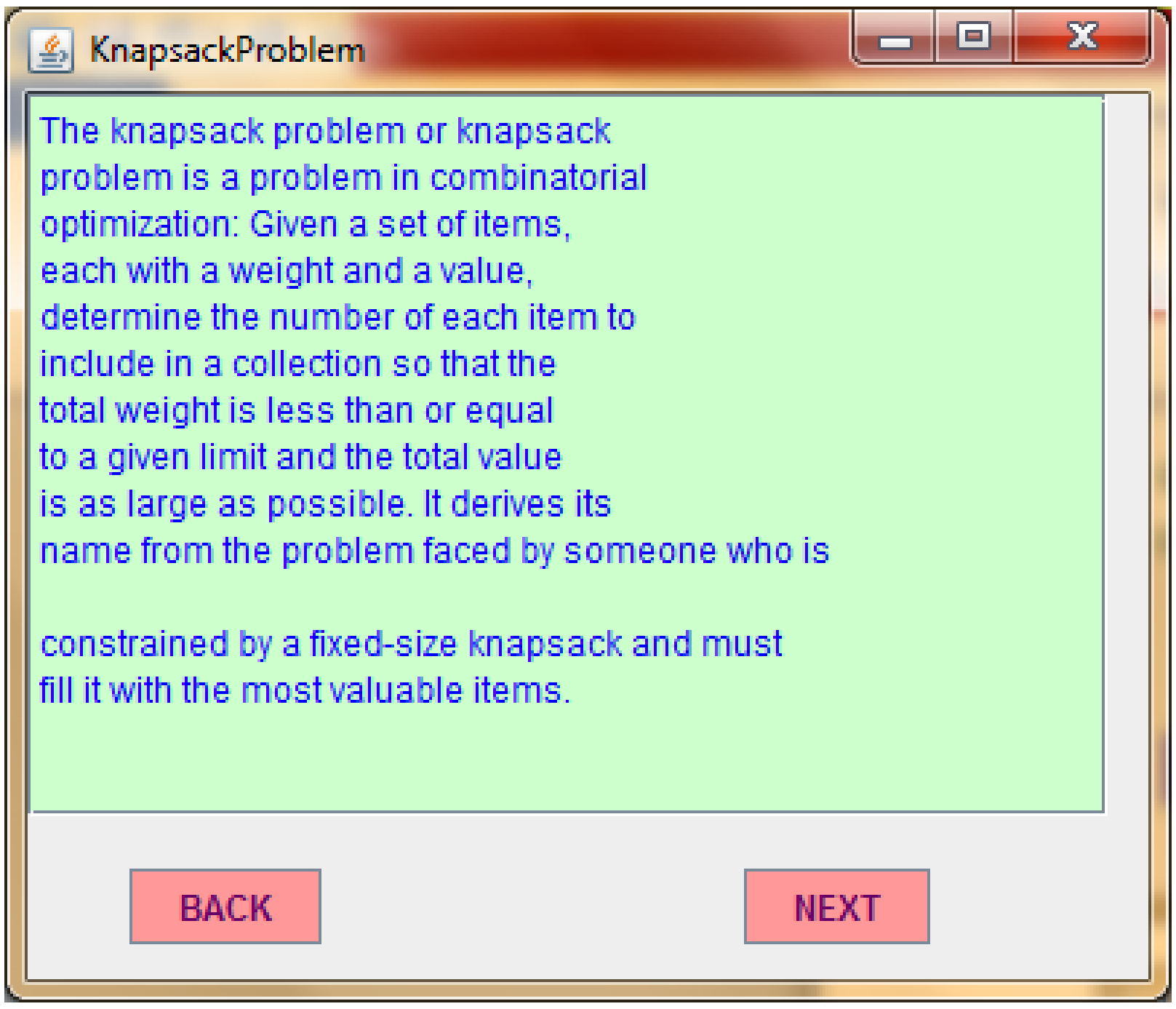

0/1 Knapsack solution window

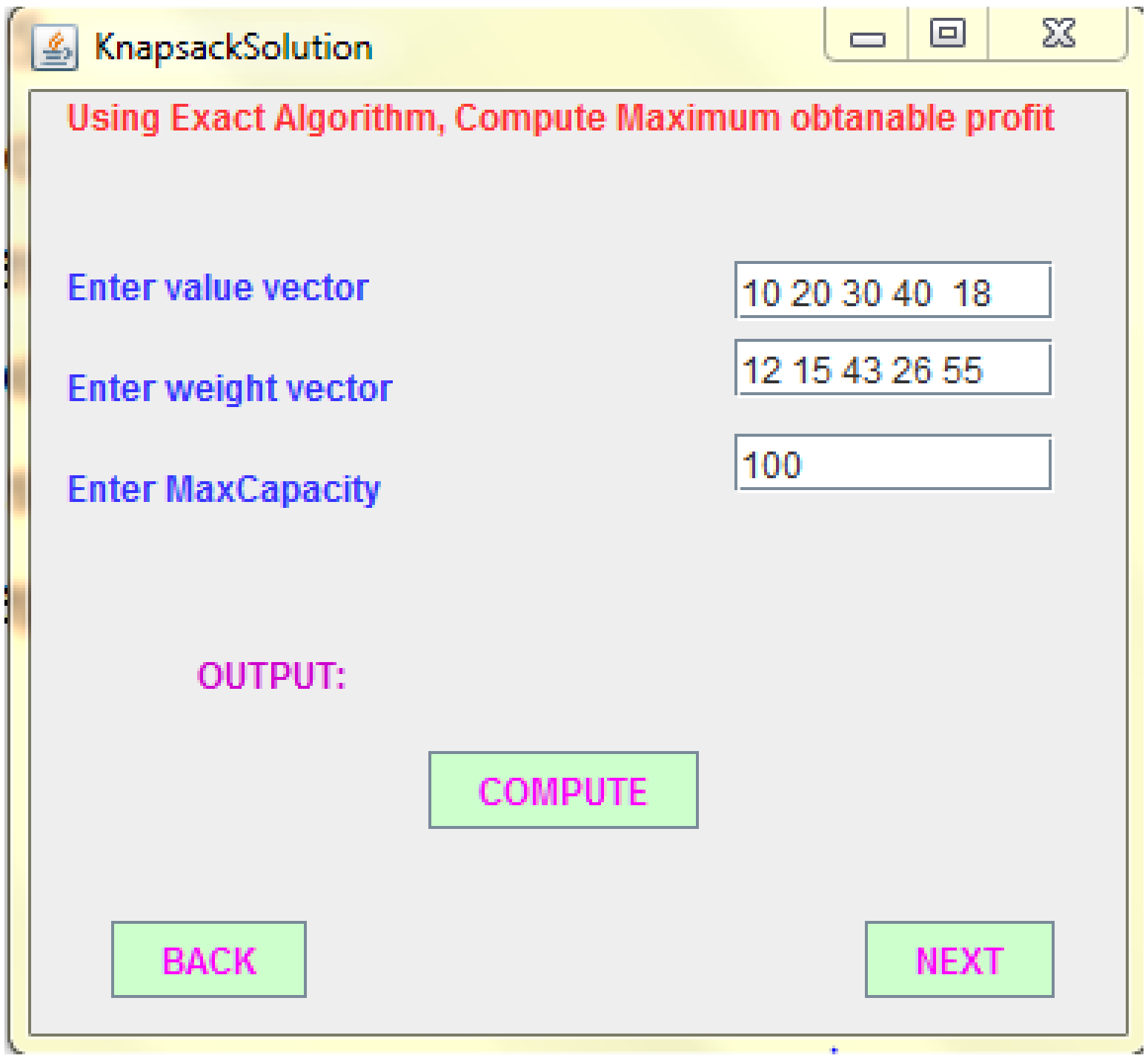

Travelling Salesman solution window

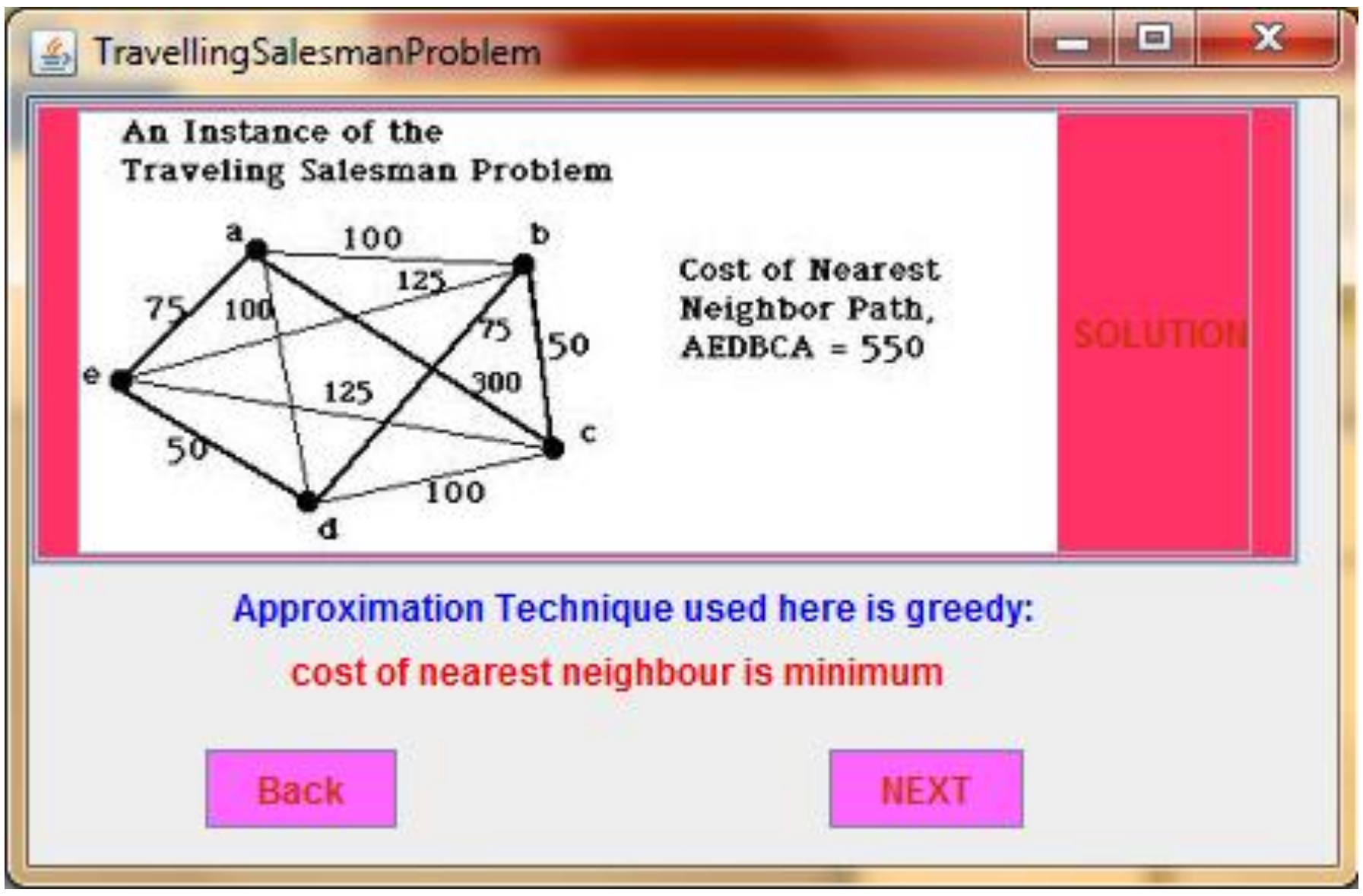

Vertex cover solution window

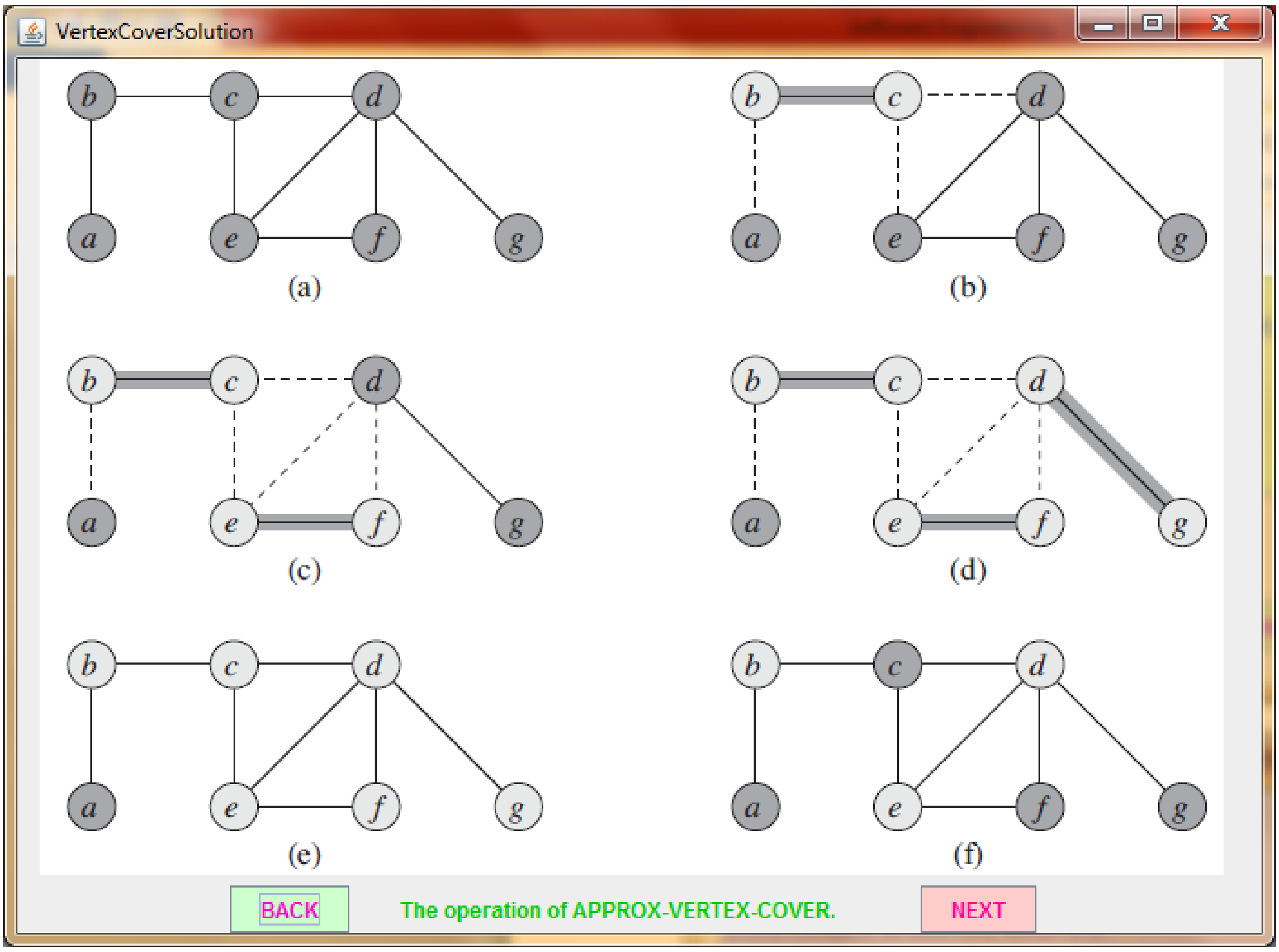

# 8. Conclusion

There are many complex subjects of discussions like the topic discussed in this project. They can be made possible to be learnt using this type of software application in a different style. This can be implemented using network or without network. In case of network application, the idea has to be extended from single user application to client server model where the application will be requested to server from client side which may be a java applet rather than an application.

# 9. Code Snippet

For every problem as well solution there is a separate class. To bring cohesion related functionality is embedded in the respective class. The Main class is like-

```
package se;

/**
 *
 * @author chiranjeev
 */
public class Main {
    /**
     * @param args the command line arguments
     */
    public static void main(String[] args) {
        // TODO code application logic here
        java.awt.EventQueue.invokeLater(new Runnable() {
            public void run() {
                new Tutorial().setVisible(true);
            }
        });
    }
}
```

This calls Tutorial class which then uses other classes and make environment flexible. Actual code can be found in the CD. It is about 2500 LOC.